\documentstyle[preprint,revtex]{aps}                   
\begin{document}
\draft
\pagestyle{empty}                                      
\null\vskip -.5in
\centerline{hep-ph/9302267 \hfill  CERN--TH.6795/93}   \vskip -.3cm
\centerline{\hfill                 NUHEP--TH--92--26}  \vskip -.3cm
\centerline{\hfill                 NTUTH--92--24}
%
\begin{title}
Two--loop Contributions of Flavor Changing Neutral Higgs Bosons\\
 to $\mu \to e\gamma$
\end{title}
\author{$^{a}$D. Chang,  $^{b}$W. S. Hou, and $^{c,d}$W.--Y. Keung}
\begin{instit}
$^{a}$Department of Physics and Astronomy,\\
Northwestern University, Evanston, Illinois 60208

$^{b}$Department of Physics, National Taiwan University,
Taipei, Taiwan 10764, R.O.C.

$^{c}$Physics Department, University of Illinois at Chicago,
Illinois 60680

$^{d}$Theory Division, CERN, CH--1211, Geneva 23, Switzerland


\end{instit}
%
\vskip -1cm
\begin{abstract}
The two--loop mechanism of Bjorken and Weinberg is used to constrain
flavor changing neutral Higgs bosons. We calculate the complete set of
two--loop diagrams for the rare decay $\mu \rightarrow e + \gamma$
induced by such neutral Higgs bosons, for arbitrary Higgs and top
masses. The analytic result is used to set limits on Higgs masses for
some recent models with specific ansatz about the flavor changing
couplings. For example, in the Cheng--Sher scenario of multi-Higgs
doublet models, all neutral Higgs bosons possess flavor changing ($f_i
\leftrightarrow  f_j)$ couplings proportional to $\sqrt{m_i m_j}$. We
find that the present limit on $\mu\to e\gamma$ implies that, in such
scheme, these neutral Higgs bosons should be heavier than $200$ GeV.
\end{abstract}
%
\narrowtext
\pagestyle{plain}

\section{Introduction}

In most theories beyond the Standard Model,
neutral Higgs boson couplings are typically
flavor changing unless special arrangements,
such as imposing discrete symmetries,
are made to eliminate them.
The mass of such Higgs bosons can be estimated
from its potential contribution to the
$K_L-K_S$ mass difference, $\Delta m_K$.
If one assumes
that the flavor changing $sdH$ vertex
has the same Yukawa coupling as that of the heaviest quark of
the same type, that is, the $b$-quark in this case, then $\Delta m_K$
implies that the Higgs mass should be at least $150$ TeV \cite{McWLS}.
This value is
much larger than the electroweak symmetry breaking scale and
immediately poses two potential problems:
(1) the existence of an unnatural hierarchy in scale;
(2) the Higgs sector would be strongly coupled and
    predictive power is lost.

To avoid such pitfalls, two options are often exercised.
The first is to impose some discrete symmetry
to achieve what is called ``natural flavor conservation (NFC)" \cite{GW},
that is, to avoid tree level flavor changing neutral currents
or couplings (FCNC). This is easily done by requiring that
only one Higgs boson vacuum expectation value (VEV)
contributes to each type of fermion mass \cite{Hguide}.
The second is to use some scheme such that tree level flavor changing
neutral Higgs couplings $Y_{ij}$ are naturally suppressed in
low energy processes.
The latter has received revived attention in the
literature recently \cite{chengsher,AHR}, in part because it may have
interesting consequences at high energies,
such as the decay \cite{hou}  of the top quark into the charm quark and a
light neutral Higgs boson.

For example, motivated by the Fritzsch ansatz \cite{Fritzsch}
of mass-mixing relations, Cheng and Sher pointed out \cite{chengsher}
that low energy FCNC constraints
may in fact be evaded in multi-Higgs doublet models {\it without}
invoking the NFC condition.
Let the contribution of the $k$-th Higgs doublet to the
fermion mass matrix be $M_{ij}^{(k)}$,
it is not unreasonable to assume that
$M_{ij}^{(k)} = X^{(k)}_{ij}\ \sqrt{m_i m_j}$ for every $k$,
where $X^{(k)}_{ij}$ is of order unity.
Upon diagonalization, the fermion mass and mixing patterns can be
roughly accounted for, but in general, neutral Higgs bosons
in mass basis would all have flavor changing couplings.
Their scheme can be summarized
by assuming that the flavor changing couplings
$\bar f_i (F^{aL}_{ij} P_L+ F^{aR}_{ij} P_R) f_j\ H_a$
of the $a$'th
neutral Higgs scalar $H_a$ have the following
natural pattern
\begin{equation}
F^{aL,R}_{ij} = {g\over {2}} {m^{aL,R}_{ij}\over M_W}
        = {g\over {2}} \Delta^{aL,R}_{ij} {\sqrt{m_i m_j}\over M_W},
\label{eq:Yukawa}
\end{equation}
with $\Delta^{aL,R}_{ij}$ of order one.
We shall refer to the lightest such scalar boson as $H_1$.
Since low energy constraints typically involve lower generation
fermions, they are evaded by the associated tiny Yukawa couplings in
Eq.(\ref{eq:Yukawa}).
For example, taking $\Delta^{1L}_{sd} = \pm \Delta^{1R}_{sd} \sim 1$,
the $\Delta m_K$
constraint is weakened. If the lightest Higgs has pseudoscalar couplings,
it is only required to be heavier than about $1$ TeV, which is roughly
the scale where the Higgs sector is expected to become strongly coupled.
For scalar Higgs couplings, the bound is further
lowered to roughly the symmetry breaking scale \cite{sheryuan}.

The Cheng-Sher scenario was not widely appreciated, and subsequent work
\cite{sheryuan} concentrated on rare decays of $\tau$, $B$ and $D$, as
well as the utility of the $\mu\to e\gamma$ process as a constraint. It
was recently realized, however, that the progressive nature of FCNC
Higgs couplings could have interesting implications for {\it very heavy}
quarks \cite{hou}. In particular, the top quark may posses non-trivial
couplings to a neutral Higgs boson $H_1$ and a charm quark, of order
$m_{ct}/v \sim \sqrt{m_c m_t}/v$. This may result in an appreciable
branching fraction for a new channel $t\to c H_1$ in the decay of the
top quark. Alternatively, neutral Higgs bosons may have appreciable
rates into $t\bar c$ type of final states. Because of the importance of
top and Higgs physics, it is natural to ask whether we have exhausted
low energy constraints on the Cheng-Sher scenario.

In this paper, we give a careful analysis of the effect of
these flavor changing neutral Higgs bosons on
the $\mu\to e\gamma$ process up to the two--loop level.
We parametrize our calculation in a way that is as model independent
as possible so that our result can be applicable to any models with flavor
changing neutral Higgs bosons.
Our one--loop result differs from previous calculation\cite{sheryuan}.
Our two--loop study not only improves on
previous rough estimates\cite{bjw,barr}, but also uncovers some
interesting characteristics that were overlooked before.
In particular, the two--loop contribution can be larger than
the one--loop result, as argued some time ago by Bjorken and Weinberg,
and the heavy--Higgs--boson effect is not decoupled.
The contributions of the two--loop graphs that contain
a heavy gauge boson loop diverge logarithmically as
the Higgs mass is taken to infinity.
However, due to a unitarity condition associated with
flavor changing neutral Higgs bosons,
if all such Higgs bosons are degenerate in mass,
their contributions mutually cancel in such diagrams.
This cancellation mechanism introduces model-dependence,
reducing one's capability to make strict predictions.
We present our results such that anyone can extract constraints
on his favorite model.
In this paper, We do not consider the charged Higgs boson,
which  gives an independent contribution with its own parameters.

\section{Bjorken--Weinberg Mechanism}

One flavor changing mode of particular interest is the celebrated
$\mu\to e\gamma$ mode. The existing limit,
at $4.9\times 10^{-11}$ \cite{PDG}, is one of the most impressive.
It cannot occur at tree level,
and it involves lepton number violation.
In the Cheng--Sher scenario,
the leading one--loop contribution has the neutral scalar
and the $\tau$ in the loop, with the photon radiated from
the internal $\tau$ line.
The usual $\mu$--$\mu$--$e$ sequence is much more suppressed.
The $\tau$ contribution (Fig.~1) to the 1--loop
branching fraction of $\mu\rightarrow e \gamma$ is
\begin{equation}
\hbox{BR}_{\hbox{1--loop}} (\mu\to e\gamma) =
      {3\over 4} {\alpha\over\pi}{m_e\over m_\mu}
         \biggl\vert\sum_a \Delta^a_{e\tau}\Delta^a_{\mu\tau}
                   {m_\tau^2\over M_{H_a}^2}
          \Bigl( \ln{m_\tau^2\over M_{H_a}^2} +{3\over 2} \Bigr)
         \biggr\vert^2
\;,
\end{equation}
where $M_{H_a}$ stands for the the mass of the $a$'th flavor changing
Higgs boson $H_a$. Clearly the contribution
from the lightest neutral scalar boson dominates in general.
This result holds if the flavor changing couplings
are purely scalar ($\Delta^{aL}_{ij} = \Delta^{aR}_{ij}$)
or purely pseudoscalar ($\Delta^{aL}_{ij} = - \Delta^{aR}_{ij}$).
Also, our result is quite
different from previous estimates \cite{sheryuan,bjw}.
Assuming lightest scalar dominance, we have
\begin{equation}
\hbox{BR}_{\hbox{1--loop}} (\mu\to e\gamma) =
        5\times 10^{-11}
      \bigl\vert \Delta^1_{e\tau}\Delta^1_{\mu\tau} \bigr\vert^2
         \Bigl({91 \hbox{GeV} \over M_{H_1} } \Bigr)^4
         \Bigl(1-0.31\ln {91 \hbox{GeV} \over M_{H_1} } \Bigr)^2 \;,
\label{BFtau}
\end{equation}
which implies that, if one takes $\Delta^1_{e\tau}\Delta^1_{\mu\tau}=1$
as in Ref.\cite{sheryuan} for the lightest scalar, any mass above 91
GeV is still phenomenologically allowed.
This limit is more stringent than that obtained in Ref.\cite{sheryuan}
because of the extra large $\ln(m_\tau^2/m_{H_1}^2)$ term.
The strong $M_{H_1}$ dependence
in Eq.(\ref{BFtau}) means that the bound on $M_{H_1}$ will improve
rather slowly with improvements on the $\hbox{BR}(\mu\to e\gamma)$ limit.

Here we wish to draw attention to an observation \cite{bjw} made by
Bjorken and Weinberg 15 years ago, that certain two--loop graphs
may in fact dominate over the one--loop contribution.
The mechanism is as follows.
Dipole transitions demand a chirality flip between the initial state and
the final state of the fermion.
For the one--loop graph involving virtual scalars,
{\it three} chirality flips are involved: twice in the
Yukawa couplings, and once in the fermion propagator.
This fact is indeed just an accident at the one--loop level,
but can be avoided at higher orders.
Clearly, at the one--loop level, the $\mu$--$\mu$--$e$ sequence
is extremely suppressed,
while even for the $\mu$--$\tau$--$e$ sequence discussed here,
one pays the price of a suppression factor
$\sqrt{m_e m_\mu} m_\tau^2/v^3 = {\cal O}(10^{-9})$.
Going to two--loop order, one pays the typical price
of $g^2/16\pi^2$, but if one could avoid two of the
extra chirality flips, one may still gain enormously against
the one--loop graph.
In a set of two--loop graphs found by
Bjorken and Weinberg,  the virtual scalar boson couples
only once to the lepton line, inducing the needed chirality flip.
Through some heavy--particle ({\it e.g.} $W$ or top) loop,
the boson is then converted into two photons,
one of which is reabsorbed by the lepton line.
Assuming  $M_H^2 \ll M_W^2$,
Bjorken and Weinberg estimated the branching fraction due to
the leading two--loop graph from the $W$ contribution (Fig.~2a) to be
\begin{eqnarray}
    \hbox{BR}_{\rm BjW}(\mu\to e\gamma)\
&\simeq&
    \frac{147}{16} \left(\frac{\alpha}{\pi}\right)^3
       {m_e \over m_\mu}
   \Biggl\vert \sum_{a>1}
                 \cos\phi_a \Delta^a_{e\mu}
                 \ln \frac{M_{H_a}^2}{M_{H_1}^2}
   \Biggr\vert^2 \;,\nonumber\\
&\simeq&
   56 \times 10^{-11}
   \Biggl\vert \sum_{a>1}
                 \cos\phi_a \Delta^a_{e\mu}
                 \ln \frac{M_{H_a}^2}{M_{H_1}^2}
   \Biggr\vert^2   \;.
\label{eq:BjW}
\end{eqnarray}
Here the neutral Higgs boson $H_a$
couples to the $W$ boson at a relative strength $\cos\phi_a$
with respect to that of the Higgs boson in the Standard Model.
It is reasonable to assume that $\cos\phi_a$ is of order one,
if the neutral Higgs boson originated from a Higgs multiplet
that contributes significantly to $SU_L(2)$ breaking.
Since this estimate works only when $M_H \ll M_W$, the present
experimental limit implies that the Cheng-Sher scenario is probably not
viable for light Higgs bosons, unless $\Delta^a_{e \mu}$ is significantly
smaller than one, or the amplitudes from different Higgs bosons cancel
each other by accident, which could happen when the relevant Higgs
bosons are degenerate in mass as discussed before.

In the context of studying the electric dipole moment ({\it edm})
of the electron within
neutral Higgs models of CP violation, Barr and Zee \cite{barrzee}
made independent observations that are analogous to that of
Bjorken and Weinberg.
Without knowing the work of Cheng and Sher,
recently Barr \cite{barr} estimated the two--loop contributions
to $\mu\to e\gamma$ for the case of very heavy Higgs bosons.
Assuming only the $W$ loop in the effective
$H\gamma\gamma$ coupling and
assuming $M_H^2 \gg M_W^2$, Barr estimated the branching fraction
from the two--loop effect to be
\begin{equation}
\hbox{BR}_{\hbox{Barr}}(\mu\to e\gamma)\ \simeq
       {3\over4}
        \biggl({\alpha\over\pi }\biggr)^3
        \biggl({ 35   \over 8  }\biggr)^2
            (\cos\phi_1\Delta^1_{e\mu})^2
       {m_e\over m_\mu} \Bigl(   {M_W\over M_{H_1}}\Bigr)^4
       \Biggl\vert \ln {M_W^2\over M_{H_1}^2}\Biggr\vert^4  \;.
\label{eq:Barr}
\end{equation}
Note that there is $M_H^{-4}$ suppression as in the one--loop case.
Taking this result seriously and assuming
$\cos\phi_1\Delta^1_{e\mu} \simeq 1$
as before, one find that the present experimental limit requires
$M_H
{\lower2mm\hbox{$\,\stackrel{\textstyle >}{\sim}\, $}}
730$ GeV in the Cheng-Sher scenario.

Given the two estimates quoted above, where the lightest
Higgs boson with flavor changing couplings is either very
light or very heavy compared to $M_W$, one may naturally be
curious about the situation for Higgs masses in between.
In this article, we report on a detailed calculation \cite{changhou} of
the complete set of two--loop diagrams where
Higgs and top masses are kept arbitrary.

\section{Complete Two--Loop Results}

The diagrams needed for the calculation of the transition dipole moment
in our case are analogous to those for
the electric dipole moments of quarks \cite{cky1} and
electrons \cite{barrzee,cky2,leigh}.
The most detailed depiction of these graphs can be found in
Ref.\cite{leigh}, where one of the external electrons should be replaced
by a muon.

The two--loop graphs of interest can be classified into three sets, A, B
and C, that are separately gauge invariant. Set A contains a heavy
fermion loop. The fermion in the loop is most likely the heaviest one
which is the top quark. They can be further classified into two gauge
invariant subsets.  The first one involves an internal photon line in
Fig.~2a, while the second is obtained by replacing the internal photon
line with a $Z$ boson line.
According to Furry's theorem, only the vector coupling of $Z$ boson
contributes to the fermion loop.  On the other hand, for both electron
electric and magnetic dipole moments, the corresponding operators of the
moments, $\sigma_{\mu\nu}$ and $\sigma_{\mu\nu}\gamma_5$, are odd under
charge conjugation, $C$, just as the vector coupling of $Z$.  Therefore
it is not hard to see that only the vector coupling of $Z$, not the
axial one, to the external leptons can contribute to these operators.
This argument can be carried  over diagrammatically to the case of the
transitional moments in the process $\mu\to e\gamma$. Since the vector
coupling of the $Z$ boson is known to be relatively suppressed, the
second group of diagrams can be ignored in the first approximation.

Set B contains a $W$ boson (and associated unphysical scalar) loop, and
can be further divided into two gauge invariant subsets. For the first
group, B$_I$, the $W$ boson loop induces a $H\gamma\gamma$ vertex as in
Fig.~2. For the second group, B$_{II}$, the internal photon line is
replaced by a $Z$ line. Just like set A, set B$_{II}$ is suppressed
compared to set B$_I$ because of the small vector coupling of $Z$ to
charged leptons. In addition, if one assumes CP invariance, then since
only the scalar components of the Higgs bosons couple to the $W$ boson
at the tree level, one expects that only the CP--even Higgs bosons will
be relevant in this case. In general, from experience with the analysis
of electric dipole moments in Refs. \cite{barrzee,cky1,cky2,leigh}, one
expects set B to dominate over set A also.

Set C involves graphs that have a different topological structure. They
can be further divided into two gauge invariant groups C$_I$ and
C$_{II}$. They correspond to graphs without or with a $Z$ boson line as
in Figs. 3, 4 of Ref.\cite{leigh}, respectively. Again, the second group
is small compared to the first due to the small $Z$ coupling.  Numerical
results of Ref.\cite{leigh} indicate that, for the case of {\it edm},
the contribution of set C is in general much smaller than sets A and B.
This conclusion should also be applicable to the transition dipole
moment.

We shall consider sets A and B first. For flavor changing leptonic
processes, the internal gauge boson line can be either the photon or $Z$
boson.  However, if one is interested in flavor changing processes
involving light quarks, a similar graph with both gauge bosons replaced
by gluons ({\it i.e.} $Hgg$ rather than $H\gamma\gamma$ or $H\gamma Z$)
can also be important.

The calculational strategy is to first calculate the one--loop effective
vertex with one gauge boson, one photon and a neutral Higgs boson in the
external lines. This has been done many times before \cite{Cahn}, and a
recent calculation can be found in Ref.\cite{barroso}. 
pseudoscalar amplitudes 
Higgs--gluon--gluon vertices 
Higgs boson are assumed to be on shell in Ref.\cite{barroso}, it is easy
to extract the result with off-shell Higgs boson as long as one can tell
which factor of Higgs mass comes from the loop momentum and which one is
due to the vertex.  The result of Ref.\cite{barroso} is consistent with
the recent calculation of electron {\it edm} \cite{leigh}, where the
Higgs boson was kept explicitly off--shell, but only $H\gamma\gamma$
contribution was given.

Here we shall concentrate on leptonic FCNC.
In that case the graphs with internal $Z$ boson are suppressed
relative to the ones with internal photon line by a factor of
$(1-4\sin^2\theta_W)/4 \sin^2\theta_W$,
which is about $0.087$ for $\sin^2\theta_W=0.23$.
Therefore one could ignore these contributions even
though they can be easily incorporated into the analysis.

We shall parametrize the relevant couplings as
\begin{eqnarray} {\cal L} =
   &-&{m_t\over v}\ \bar t(\Delta^a_{tt} P_L+{\Delta^a_{tt}}^* P_R) t\ H_a
\nonumber\\
   &-& {\sqrt{m_{\mu} m_e}\over v}\
         \bar e (\Delta_{e\mu}^{aL} P_L+\Delta_{e\mu}^{aR} P_R) \mu\ H_a
    + gM_W\cos\phi_a\ W^+W^-H_a +\cdots \;.
\label{eq:Lagrangian}
\end{eqnarray}
Here $v=(\sqrt{2} G_F)^{-{1\over 2}} \simeq 246 \hbox{ GeV}$.
If one imposes CP conservation then $\hbox{Im}\ (\Delta^a_{tt})^2 =0$
and $\hbox{Im}\ (\Delta_{e\mu}^{aL} \Delta_{e\mu}^{aR*}) =0$.
Note also that in case of CP conservation, $\cos\phi_a$ is nonzero
only for those CP--even scalar Higgs bosons.
(For the Higgs boson in Standard Model, $\Delta_{tt} =1$)

To simplify long expressions, we define the
reduced amplitude $A$, which is dimensionless,  for the transition
$\mu\to e\gamma(\epsilon,k)$ as follows:
\begin{equation}
i{\cal M}={e {\sqrt 2} G_F \alpha \over 16\pi^3} \sqrt{m_{\mu} m_e}\
\epsilon^{\mu} k^{\alpha} \sigma_{\mu\alpha } (A_LP_L+A_RP_R)  \;,
\label{eq:reduceamp}
\end{equation}
and the branching fraction is
\begin{equation}
    \hbox{BR}(\mu\to e\gamma)
 =  \frac{3}{4} \left(\frac{\alpha}{\pi}\right)^3
       {m_e \over m_\mu}
   \Bigl( \case1/2 \Bigl\vert A_L   \Bigr\vert^2
         +\case1/2 \Bigl\vert A_R   \Bigr\vert^2  \Bigr)
 =  4.5 \times 10^{-11}
           \Bigl( \case1/2 \Bigl\vert A_L   \Bigr\vert^2
                 +\case1/2 \Bigl\vert A_R   \Bigr\vert^2  \Bigr)
\;.
\end{equation}
Note that CP conservation will require $\hbox{Im}\ (A_L A_R^*) =0$.
For set A with the top--quark loop, the $H\gamma\gamma$ or
$H\gamma Z$ vertices already contain one power of external photon
momentum. Therefore, we can set the virtual photon momentum
and the Higgs boson momentum
to be equal and the two--loop result can be easily
produced. The $H\gamma\gamma$ contribution gives
\begin{equation}
A_{t\hbox{--loop}}^{L,R\ H\gamma\gamma}(\mu \rightarrow e + \gamma)
= 3Q_t^2  \sum_a \Delta^{a\ L,R}_{e\mu} 2
     \Bigl[
                          \hbox{Re}\ \Delta^a_{tt}\ f(z_{ta})
      - i \lambda_5^{L,R} \hbox{Im}\ \Delta^a_{tt}\ g(z_{ta})
     \Bigr]  \;,
\label{eq:tloop}
\end{equation}
where $z_{ta}=m_{t}^{2}/M_{a}^{2}$.
The chirality factors are defined as $\lambda_5^L=-1$ and $\lambda_5^R=1$.
The scalar $Ht\bar t$ Yukawa coupling Re $\Delta_{tt}^a$ is
associated with the following function,
\begin{equation}
 f(z)={1\over 2}z\int_0^1 dx {1-2x(1-x) \over x(1-x)-z}
                                           \ln {x(1-x) \over z}.
\end{equation}
The pseudoscalar coupling Im $\Delta_{tt}^a$ is associated with
\begin{equation}
g(z) = {1 \over 2}z \int_0^1 dx{1 \over x(1-x)-z}\ln{x(1-x) \over z}.
\end{equation}
We have closely followed the notations of Ref.\cite{cky2}.

If CP is invariant, the $a$'th Higgs boson, when  it couples to the top quark,
is either a scalar ($\hbox{Im}\ \Delta^a_{tt}\ =0$), or a pseudoscalar
($\hbox{Re}\ \Delta^a_{tt}\ =0$). However, we do not have relations
between $\Delta^{a\ L}_{e\mu}$ and $\Delta^{a\ R}_{e\mu}$,
except that they are relatively real.

For the $Z$--mediated diagrams,
\begin{eqnarray}
A_{t\hbox{--loop}}^{L,R\ HZ\gamma}(\mu \rightarrow e + \gamma) =
{(1-4\sin^2\theta_W)(1-4Q_t\sin^2\theta_W)
\over 16\sin^2\theta_W\cos^2\theta_W}
\nonumber\\
\times
3 Q_t \sum_a \Delta^{a\ L,R}_{e\mu} 2
        \Bigl[\hbox{Re}\ \Delta^a_{tt} \tilde f(z_{ta}, z_{tZ})
                  - i\lambda_5^{L,R}
              \hbox{Im}\ \Delta^a_{tt} \tilde g(z_{ta}, z_{tZ})
        \Bigr] \;.
\end{eqnarray}
Here
$\tilde f(x,y)=yf(x)/(y-x) + xf(y)/(x-y)$ and similarly
$\tilde g(x,y)=yg(x)/(y-x) + xg(y)/(x-y)$.
We have also extended the previous definition to denote
$z_{tZ}=m_{t}^{2}/M_{Z}^{2}$.
Note that, in this Bjorken--Weinberg mechanism, there is only one power of
light quark mass suppression \cite{bjw}, which has been explicitly
written out in Eq.(\ref{eq:reduceamp}).

To derive the contribution of the bosonic loops, we shall classify the graphs
into two gauge invariant types.
The first type does not depend on Higgs mass in
their couplings while the second set does.  As a result, the first set is
power suppressed by the Higgs mass while the second set is
logarithmically increasing when the Higgs mass becomes very large,
which is a very intriguing situation.  For Higgs mass larger than a certain
value the second type dominates.  We shall only present the combined
contribution of the two types.

The $H_aWW$ vertex in Eq.(\ref{eq:Lagrangian}) is parametrized as
$g M_W g^{\mu \nu} \cos\phi_a$, where $\cos\phi_a$ is a Higgs mixing
angle. For the Standard Model Higgs boson, $\cos\phi = 1$.
Before we proceed with our results, it is important to point out
a unitarity constraint on the flavor changing neutral couplings.
One can always make linear combinations of the scalar doublet fields
such that only one doublet is responsible for symmetry breaking.
It is the scalar component of this doublet that couples to $W$ boson pairs.
Since this combination is also responsible for generating masses
to the fermions, its Yukawa couplings should be automatically flavor
conserving. Upon diagonalization of the Higgs boson mass matrix, all
neutral Higgs bosons should in general possess flavor changing
couplings. However, the above observation leads to a unitarity condition
\begin{equation}
\sum_{a} \cos\phi_a \Delta^{a\ L,R}_{ij} = 0\ (\hbox{ for}\ i \neq j),
\label{eq:unitarity}
\end{equation}
which basically reflects the fact that the scalar doublets
that mediate flavor violation must have zero vacuum expectation value
at tree level.
One important consequence is that, for the graphs in
set B, terms that are independent of Higgs mass are cancelled away.

For the $H\gamma\gamma$ case, one obtains the two--loop amplitude
\begin{equation}
A_{W\hbox{--loop}}^{L,R\ H\gamma\gamma}(\mu \rightarrow e + \gamma) =
-\sum_a \cos\phi_a \Delta^{a\ L,R}_{e\mu}
   \left[ 3 f(z_{a}) + 5 g(z_{a}) + \case3/4 g(z_a)+\case3/4 h(z_a)
\right] \ ,
\label{eq:Wloop}
\end{equation}
with $z_{a}=M^2_W/M^2_{H_a}$. The function $h(z)$ is defined as
\begin{equation}
h(z)=z^2 {\partial \over \partial z}\biggl({g(z) \over z}\biggr)
    ={z\over2}\int_0^1 {dx\over z-x(1-x)}
         \biggl[1+{z\over z-x(1-x)}\ln{x(1-x)\over z} \biggr].
\end{equation}

It is straightforward to see that $A^{L,R} \propto \Delta^{a\ L,R}_{e\mu}$
for this $W$--loop amplitude and other purely bosonic loop contributions,
unlike
the situation for the $t$--loop with the pseudoscalar coupling
Im$\Delta_{tt}^a$. Therefore we shall drop
the chirality label for amplitudes from sets B and C for brevity.
Note that if CP is conserved, $\cos\phi_a = 0$ unless the $H_a$ is a scalar.
Therefore the flavor changing pseudoscalar Higgs boson does not have
contribution of this class in such case.

It is useful to know the shapes of these functions $f$, $g$ and $h$.
Numerically, $f(1)$ is about $0.8$, while $g(1)$ is about $1.2$.
The general $z$ dependence of these functions are given in Fig.~3.
Unless $z$ is very small or very large, these functions are of order unity.
For very large or very small  $z$ \cite{barrzee,cky1,cky2},
\begin{eqnarray}
 f(z\gg 1) \sim \case1/3 \ln z + \case{13}/{18},   \quad
&g(z\gg 1) \sim \case1/2 \ln z + 1             ,   \quad
&h(z\gg 1) \sim -\case1/2(\ln z + 1)           ,   \nonumber\\
 f(z\ll 1) \sim \case{z}/2 (\ln z)^2           ,   \quad
&g(z\ll 1) \sim \case{z}/2 (\ln z)^2           ,   \quad
&h(z\ll 1) \sim            z\ln z              .
\end{eqnarray}
For the large $z$ asymptotic forms, we obtain
Eq.(\ref{eq:BjW}) in the light--Higgs limit
from Eq.(\ref{eq:Wloop}). It is tempting to use also Eq.(\ref{eq:Wloop})
to find the heavy--Higgs $z \ll 1$ limit,  which will produce the estimate
Eq.(\ref{eq:Barr}) given in ref.\cite{barr}.
However, this estimate clearly overlooks other
non--decoupling contributions in Fig. 2c,d that we will discuss.

For the $HZ\gamma$ case, one has
\begin{eqnarray}
A_{W\hbox{--loop}}^{HZ\gamma}(\mu \rightarrow e + \gamma)&=& -
{1-4\sin^2\theta_W \over 4\sin^2\theta_W}
\sum_a \cos\phi_a \Delta^a_{e\mu}
\nonumber\\
&\times& \Bigl[
   \case1/2 (5- \tan^2\theta_W) \tilde f(z_{a},z_Z)
  +\case1/2 (7-3\tan^2\theta_W) \tilde g(z_{a},z_Z)
\nonumber\\
&\ &  +\case3/4 g(z_{a})
      +\case3/4 h(z_{a})
\Bigr]
\; .
\label{eq:WloopZ}
\end{eqnarray}
with $z_Z=M^2_W/M^2_Z$.
The suppression factor of $(1-4\sin^2\theta_W)/ 4\sin^2\theta_W$ comes
from the vector part of the $eeZ$ coupling.
The $W$--loop contribution of  $HZ\gamma$ is about 10\% of
that of $H\gamma\gamma$ and they have the same sign.

If we assume that CP is conserved and the Higgs boson is
a CP--odd pseudoscalar, it does not couple to the $W$ boson
and there will be no contribution from set B and C.
However, there will be Im$\Delta_{tt}^a$
contributions in Eq.(\ref{eq:tloop}) from set A.

The type of diagrams that involve the Higgs mass squared in
the coupling are shown in Fig.~2b,c,d, which are Fig.~2k,l,m in
Ref.\cite{leigh} respectively.
They contain the coupling of the physical Higgs boson $H$
to the unphysical Higgs pair $G^+G^-$.
An important exception is the contribution related to Fig.~2b.  This diagram
has been grouped into Fig.~2a with the bosonic inner loop because they
are gauge related.  The contribution proportional the
Higgs mass squared, from the $HG^+G^-$ vertex, can be combined
with part of Fig.~2a to form a vertex that is proportional to the inverse
Higgs propagator which then cancels with the Higgs propagator
in the outer loop.
Thus the resulting contribution for each $H_a$ is independent of $M_{H_a}$
and therefore cancels each other completely because
of the unitarity condition Eq.(\ref{eq:unitarity}).

The contribution from Fig.~2c,d gives
\begin{equation}
A_{G\hbox{--loop}}^{H\gamma\gamma}(\mu \rightarrow e + \gamma) = -
\sum_a \cos\phi_a \Delta^a_{e\mu}
{1 \over 2z_{a}} \Bigl[  f(z_{a})-g(z_{a}) \Bigr],
\label{eq:Gloop}
\end{equation}
and
\begin{eqnarray}
A_{G\hbox{--loop}}^{HZ\gamma}(\mu \rightarrow e + \gamma) = -
{1-4\sin^2\theta_W \over 8\sin^2\theta_W} (1-\tan^2\theta_W)
\nonumber\\
\times \sum_a \cos\phi_a \Delta^a_{e\mu}
{1 \over 2 z_a}
   \Bigl[ \tilde f(z_a,z_Z) - \tilde g(z_a,z_Z) \Bigr]  \ .
\label{eq:GloopZ}
\end{eqnarray}
Note that $f(1) - g(1) = -0.4$, while for small $z$,
which corresponds to the case of very large Higgs mass,
$f(z) - g(z) \sim z({\ln}z + 2)$.
This leads to the peculiar situation where the contribution
increases logarithmically with the Higgs boson mass.
The coefficients are small enough that these contributions
are not so significant as compared to the $W$--loop contribution
discussed earlier, except for the case of very heavy Higgs boson.
Of course, one may not trust the perturbative estimate if the Higgs mass
becomes too heavy and the Higgs self-coupling becomes nonperturbative.

The contribution of the two--loop graphs in set C can be easily translated
from the calculation of Ref.\cite{leigh}.  The result is
\begin{eqnarray}
A_{\rm C}(\mu \rightarrow e + \gamma) =
- {1 \over 4 \sin^2\theta_W}
\sum_a \cos\phi_a \Delta^a_{e\mu}
\Bigl[ D_e^{(3a)}(z_a) + D_e^{(3b)}(z_a) + D_e^{(3c)}(z_a)
\nonumber\\
+D_e^{(3d)}(z_a) + D_e^{(3e)}(z_a) \
+ D_e^{(4a)}(z_{Za}) + D_e^{(4b)}(z_{Za}) + D_e^{(4c)}(z_{Za})
\Bigr],
\end{eqnarray}
%
where the functions $D_e^{(3a,b,c,d,e)}(z)$ and $D_e^{(4a,c)}(z)$
are given in Appendix B of Ref.\cite{leigh},
$z_{Za} = M_Z^2/M_{H_a}^2$ and
$D_e^{(4b)}(z) = 4 \sin^2\theta_W \tan^2\theta_W D_e^{(3c)}(z)$.
Note that the terms with functions $D_e^{(3a,b,c,d,e)}(z_H)$ belong to
the group $C_I$, while the rest belong to the second group $C_{II}$.
As commented
earlier the second group is suppressed relative to the first group.
The reason one can easily translate the calculation of
{\it edm} from these graphs into contributions to the transitional
magnetic moment is because the Higgs line is always attached
to one of the external fermion lines,
and because the Higgs boson only has scalar
couplings to gauge bosons at tree level.   Therefore, in the case of
{\it edm} the Higgs coupling to fermions is always pseudoscalar,
while its coupling for transitional magnetic moment is always scalar.

Just as the calculation of {\it edm} in Ref.\cite{leigh},
the contribution of the graphs in
set C to the transitional magnetic moment is also small.  We therefore do
not include them in our numerical analysis.

\section{Discussion and Conclusion}

In our numerical analysis, we shall ignore CP violation and
take $\Delta^a_{tt}$ to be real. We also assume that
$\Delta_{e\mu}^{aL} = \Delta_{e\mu}^{aR} =\Delta_{e\mu}^a$,
{\it i.e.}, scalar Higgs couplings. Under this condition, the reduced
amplitudes are simplified $A^L=A^R=A$.
Comparison with experiment is given in Fig.~4.
The result in general depends on three parameters, in addition to the
unknown top and Higgs masses. They are the parameters
$\Delta_{tt}$ and $\Delta_{e\mu}$ (Eq.(\ref{eq:Lagrangian}))
which parametrize the $ttH$ and $e\mu H$ couplings,
and $\cos\phi_a$, which parametrizes the $H_aWW$ coupling.
We set these parameters to one in our figures as a reference point.
At this moment, we pretend that the contributions from different Higgs bosons
do not strongly cancel each other. The results
due to the contributions from a single Higgs boson
are shown in Fig.~4.
Numerically the contribution from the top--quark loop is generally smaller
than that from the $W$--boson loop.
Also, the contributions due to the $Z$
boson can be ignored although we have included them in our
numerical analysis.

It is instructive to look at the numerical results at the amplitude
level. In Fig.~5, we show the reduced amplitudes $A$'s
from various sources. The error bar of
the data point indicates the experimental bound on $\mu \to e\gamma$,
$\vert A \vert
{\lower2mm\hbox{$\,\stackrel{\textstyle <}{\sim}\, $}}
1$.
The data point is used to guide the reader's eyes. It is purposely
located at the lower bound of the Higgs mass at 91 GeV from the
one--loop result.
Because the two--loop $W$ contribution does not vanish in the heavy--Higgs
limit, one may need to sum up contributions from different Higgs bosons.
For the case of the two--doublet model, the unitarity condition implies that
the $W$ amplitude is just the difference between those from the Higgs
bosons at two separate masses. It is understood that when the Higgs
bosons are degenerate in mass, the $W$--loop contributions cancel each
other.

Some interesting features deserve special attention:\\
(1) As $M_H \rightarrow \infty$, the amplitude in Eq.(\ref{eq:Gloop})
does not go to zero, unlike Barr's estimate in Eq.(\ref{eq:Barr}).
In fact it goes to infinity as $\ln M_H$.
This non--decoupling behavior is curious but can be easily understood.
In Feynman gauge, the non--decoupling graphs involve
neutral Higgs coupling to unphysical charged scalars,
which is proportional to Higgs mass squared.
Such couplings are dictated by the gauge symmetry and
its breaking.  At tree level, the Higgs field that appears
in the $W^+W^-H$ coupling is the scalar component of the Higgs doublet
that generated the symmetry breaking (the other three components
are precisely the unphysical scalar bosons).
This component is in general not a mass eigenstate of course.
It can be expressed as the sum of the scalar components of the Higgs fields
that participate in the breaking of $SU_L(2)$,
with coefficients proportional to their respective VEV's.
Therefore any Higgs boson that couples to the $W$ pair has to originate
from some weak multiplet that contributes significantly to the breaking
of $SU_L(2)$.  Hence the natural scale for such Higgs bosons should be
the $SU_L(2)$ breaking scale, $v$.  If their masses are artificially
pushed much higher, say, by fine tuning, their physical consequences
would {\it not} decouple. For similar reasons one can also conclude
\cite{chk} that, in Standard Model, the Higgs contribution to $g-2$ of
charged leptons would also not decouple.  For very large Higgs mass, it
should diverge as $\ln M_H$.\\
(2) In set B, the contribution from Eqs.(\ref{eq:Gloop},\ref{eq:GloopZ})
that gives rise to the non--decoupling behavior has a different
sign compared to the other part from
Eqs.(\ref{eq:Wloop},\ref{eq:WloopZ}).
At a low Higgs mass below $200$ GeV, the non--decoupling contribution
are small. This is the region where Barr's
estimate applies \cite{barr}. However, around $600$ GeV
a perfect cancellation occurs as shown in the solid curve Fig.~5.
Even if the contribution from the top--quark loop
is included, it will only shift the position of
the amplitude zero by a small amount, depending on the value of $m_t$.
If the mass of the flavor changing neutral Higgs boson
happens to lie in this region, $\hbox{BR}(\mu\to e\gamma)$ would be
very suppressed and further improvements of the
experimental limits will not be very constraining.
Such cancellation behavior can potentially become
a crucial issue in the future.\\
(3) The present experimental limit requires
$M_H\
{\lower2mm\hbox{$\,\stackrel{\textstyle >}{\sim}\, $}}
200$ GeV
under the same simplifying assumptions made earlier. This is a
factor of at least 2 better than the one--loop limit. Our conservative
limit is substantially lower than the bound 730 GeV from
Eq.(\ref{eq:Barr}), where only some of the $W$--loop contributions are
included. It turns out that other contributions from the non--decoupling
term and the $t$--loop diagram reduce the bound substantially.

(4) The result is only mildly sensitive to the top quark mass.
However, there exist models in which the flavor changing neutral Higgs
bosons have very small couplings to the $W$ boson.
That is, $\cos\phi$ may be very small.
In that case the top quark contribution dominates at two--loop and
is still larger than the one--loop result.
Conversely, there are some other models in which
the Higgs boson that couples to leptons is different from
the one that couples to up--type quarks.
In that case, the parameter $\Delta_{tt}$
would be zero and the top--loop contribution should be ignored.\\
(5) If the Higgs boson couples also to the down type quarks as in the general
Cheng-Sher scheme, the constraint from $K_L-K_S$ mass difference
in general would dominate over those from $\mu\to e\gamma$,
although they are modulated by different $\Delta_{ij}$ factors.\\
(6) The limit on $\mu\to e\gamma$ of course will improve in the future.
However, note that most of the severe low energy
FCNC constraints such as $K_L$-$K_S$ mass difference,
$B^0$-$\bar B^0$ mixing, $K_L\to \mu e$
and $\mu\to e\gamma$ {\it etc.},
originate from processes involving down--type quarks and charge leptons.
FCNC constraints involving up--type quarks
({\it e.g.} $D^0$-$\bar D^0$ mixing) are rather weak.
As pointed out in Ref.\cite{hou}, it is in fact easy to avoid
constraints from $K$, $B$ and $\mu$ systems completely,
by assuming NFC for down--type quarks and for charged leptons.
Although the ansatz may seem a bit artificial, it does, however,
permit tantalizing phenomenological consequences
for the top quark \cite{hou},
despite the depressed $\mu\to e\gamma$ transitions.

        To conclude, we have derived the result for the two--loop
contribution of flavor changing neutral Higgs bosons to the celebrated
rare decay $\mu \to e\gamma$, for arbitrary Higgs and top masses.
This is one of the rare situations in which higher
order contributions actually dominate over lower order ones.
The numerical consequences depend, of course, on the model.
For the generic case in the scheme of Cheng and Sher \cite{chengsher},
the result shown in Fig.~4 improves
the one--loop bound by more than a factor of two.
The curious behavior of non--decoupling of very heavy Higgs boson
effects at two--loop is emphasized.
\acknowledgments
DC and WSH wish to thank many useful discussions with S. Barr.
DC also wishes to thank L.-F. Li and
L. Wolfenstein, for bringing his attention to the paper
by Bjorken and Weinberg\cite{bjw}, and
P.B. ~Pal for discussions on related issues.
This research is funded in part by the U.S. Department of Energy.

\vskip -1cm
\figure{A one--loop Feynman diagram for $\mu \rightarrow e + \gamma$
through the $\tau$ lepton as the intermediate fermion.}
\figure{Feynman diagrams for $\mu \rightarrow e + \gamma$.
The generic inner--loop in (a)  involves the $t$--quark, the $W$ boson and its
ghost. For purely bosonic contributions, diagram (a) includes
sea-gull graphs and other gauge related graphs except that we separate
out those with vertices $G^+G^-H$ $\propto M_H^2$
in different diagrams (b), (c) and (d).
We also do not show conjugate diagrams with lines of the
neutral gauge boson and the Higgs boson exchanged.
}
\vskip -1cm
\figure{Numerical values for the functions $f(z)$, $g(z)$ and $h(z)$.}
\vskip -1cm
\figure{Numerical estimate of the $\mu \rightarrow e + \gamma$. The
dash--dotted  line is the one--loop result via the intermediate $\tau$
lepton. The two--loop result,  assumming coming from one single neutral
Higgs boson at $M_H$, is given by the solid (dashed) curve for the case
$m_t=100$ (200) GeV and $\cos\phi_a\Delta_{e\mu}^a=1$,
$\Delta_{tt}=1$.}
\vskip -1cm
\figure{Reduced amplitudes $A$'s for the process $\mu \rightarrow e + \gamma$.
The dotted  line is the one--loop result via the intermediate $\tau$
lepton. The data point is located at the $M_H$ lower bound due to the
one--loop result. The error bar indicates the experimental bound on the
reduced amplitude.
The two--loop $t$ contribution is given by the dashed
(dash--dotted) curve for the case $m_t=100$ (200) GeV and
$\Delta_{tt}=1$. The two--loop $W$ contribution, for the case
$\cos\phi_a\Delta_{e\mu}^a=1$,
is given by the solid curve, which does not vanish in the heavy--Higgs
limit.
}
\end{document}